\documentclass[11pt,dvips]{article}

\usepackage{epsfig,times} 
\usepackage{picinpar}
\usepackage{wrapfig}
\usepackage{floatflt}
\makeatletter
\renewcommand{\section}{\@startsection{section}{1}{0in}
	{0.4\baselineskip}{0.1\baselineskip}{\Large\bf}}
\renewcommand{\subsection}{\@startsection{subsection}{2}{0in}
	{0.25\baselineskip}{-\baselineskip}{\large\bf}}
\renewcommand{\subsubsection}{\@startsection{subsubsection}{3}{0in}
	{0.1\baselineskip}{-\baselineskip}{\normalsize\bf}}
\makeatother
\begin{document}

%
\makeatletter\newcommand{\ps@icrc}{
\renewcommand{\@oddhead}{\slshape{OG.2.1.25}\hfil}}
\makeatother\thispagestyle{icrc}
%
%

\begin{center}
%
{\LARGE \bf TeV gamma-ray observations from nearby AGNs by Utah Seven Telescope Array}
\end{center}

%
%
\begin{center}
{\bf The Utah Seven Telescope Array collaboration}
\end{center}
\begin{center}
{\bf 	T.Yamamoto$^{1}$, 
	N.Chamoto$^{2}$, M.Chikawa$^{3}$, S.Hayashi$^{2}$, Y.Hayashi$^{4}$, N.Hayashida$^{1}$,
	K.Hibino$^{5}$, H.Hirasawa$^{1}$, K.Honda$^{6}$, N.Hotta$^{7}$,
	N.Inoue$^{8}$, F.Ishikawa$^{1}$, N.Ito$^{8}$, S.Kabe$^{9}$,
F.Kajino$^{2}$, T.Kashiwagi$^{5}$, S.kakizawa$^{19}$, S.Kawakami$^{4}$, Y.Kawasaki$^{4}$,
     N.Kawasumi$^{6}$, H.Kitamura$^{16}$, K.Kuramochi$^{11}$, E.Kusano$^{9}$,
 E.C.Loh$^{12}$, K.Mase$^{1}$, T.Matsuyama$^{4}$, K.Mizutani$^{8}$, Y.Morizane$^{3}$,
    D.Nishikawa$^{1}$, M.Nagano$^{18}$, J.Nishimura$^{13}$, T.Nishiyama$^{2}$,
	M.Nishizawa$^{14}$, T.Ouchi$^{1}$, H.Ohoka$^{1}$, M.Ohnishi$^{1}$, S.Osone$^{1}$,
	To.Saito$^{15}$, N.Sakaki$^{1}$, M.Sakata$^{2}$, M.Sasano$^{1}$,
    H.Shimodaira$^{1}$, A.Shiomi$^{8}$, P.Sokolsky$^{12}$, T.Takahashi$^{4}$,
	S.F.Taylor$^{12}$, M.Takeda$^{1}$, M.Teshima$^{1}$, R.Torii$^{1}$, M.Tsukiji$^{2}$,
	Y.Uchihori$^{16}$,  Y.Yamamoto$^{2}$, K.Yasui$^{3}$, S.Yoshida$^{1}$,
	H.Yoshii$^{17}$, and T.Yuda$^{1}$
}\\
\vspace{1.0ex}
{\it $^{1}$Institute for Cosmic Ray Research, University of Tokyo, Tokyo 188-8502, Japan\\
\it $^{2}$Department of Physics, Konan University, Kobe 658-8501, Japan\\
\it $^{3}$Department of Physics, Kinki University, Osaka 577-8502, Japan\\
\it $^{4}$Department of Physics, Osaka City University, Osaka 558-8585, Japan\\
\it $^{5}$Faculty of Engineering, Kanagawa University, Yokohama 221-8686, Japan\\
\it $^{6}$Faculty of Education, Yamanashi University, Kofu 400-8510, Japan\\
\it $^{7}$Faculty of Education, Utsunomiya University, Utsunomiya 320-8538, Japan\\
\it $^{8}$Department of Physics, Saitama University, Urawa 338-8570, Japan\\
\it $^{9}$High Energy Accelerator Research Organization (KEK), Tsukuba 305-0801, Japan\\
\it $^{10}$Department of Physics, Kobe University, Kobe 657-8501, Japan\\
\it $^{11}$Faculty of Science and Technology, Meisei University, Tokyo 191-8506, Japan\\
\it $^{12}$Department of Physics, University of Utah, Utah 84112, USA.\\
\it $^{13}$Yamagata Academy of Technology, Yamagata 993-0021, Japan\\
\it $^{14}$National Center for Science Information System, Tokyo 112-8640, Japan\\
\it $^{15}$Tokyo Metropolitan College of Aeronautical Engineering, Tokyo 116-0003, Japan\\
\it $^{16}$National Institute of Radiological Sciences, Chiba 263-8555, Japan\\
\it $^{17}$Department of Physics, Ehime University, Matsuyama 790-8577, Japan\\
\it $^{18}$Department of Applied Physics and Chemistry, Fukui University of Technology, Fukui 910-8505, Japan\\
\it $^{19}$Department of Physics, Shinshu University, Matsumoto 390-8621, Japan\\
}
\end{center}

\begin{center}
{\large \bf Abstract\\}
\end{center}
\vspace{-0.5ex}
%
%

We have investigated TeV gamma-ray emissions from nearby
X-ray selected BL Lac objects using the Utah Seven Telescope Array
for more than two years.
These objects can be considered as potential sources of TeV gamma rays 
with inverse compton model.
The gamma-ray flares from Mrk421 and Mrk501 found in these observation
will be reported.

%

\vspace{1ex}

%
%
\section{Introduction:}
\label{intro.sec}
Several sources, Active Galactic Nuclei (AGNs) and Super Nova Remnants (SNRs)
are identified as TeV gamma-ray sources so far by the ground based Cherenkov
telescopes.
TeV gamma-ray observations of AGNs give us an opportunity to investigate
the environment near the central massive black hole in the AGN and 
the particle acceleration in the jet.
In addition, detail study of the energy spectrum
of the TeV gamma rays from different distance sources makes 
it possible to measure the density of inter-galactic infrared photons. 
With these reasons, the study of TeV gamma-ray emission from AGNs is
very important.
The nearest X-ray selected BL Lacs, Mrk421 (z=0.031) and Mrk501(z=0.034) 
are identified as TeV gamma-ray sources.

Using the Utah Seven Telescope Array, we have observed 
Mrk421 and Mrk501 intensively since early 1997. 
We have started
the survey observation of XBL Lacs since the autumn of 1997 
just after the detection of flares of TeV gamma-ray emission from
Mrk501 in 1997.
Survey observations were carried out more than 700 hrs so far.
Some results from the survey observation in 1997, 1998 and 1999
are reported in this paper.

\section{Experiment:}
\label{obs.sec}
%
Table 1 shows observation time under good weather
conditions for each sources. 
We have selected nearby XBL Lacs with the redshift less than 0.2 
in the northern hemisphere (Stecker et al. 1996).
Observation targets list of the day was made every night 
by choosing the objects which pass thorough near the zenith.
According to the target list, all telescopes track 
the same target.
We adopted the tracking method called as ``Raster Scan''. 
In this mode, the tracking center of the telescopes scans 
the square region of $\pm0.5^{\circ}$
in right ascension and declination coordinate centered on the target with
a cycle of 48minutes.
After the observation with one raster scan cycle, 
the data is analyzed immediately.
If we find unusual event excess at the source position, 
the observation is continued for another raster scan cycle. 
If there is no excess, all telescopes begin to track next target 
as in schedule. 
These data are transfered to Japan through the Internet 
for detail analysis.

Optimization of the cut values for image parameters was carried out
using the data of Mrk501 obtained in 1997. 
Threshold energy and mode energy of the detector for gamma rays are
estimated as 600 GeV and 900GeV, respectively. 
By using the optimized cut values, we estimate by the Monte Carlo simulation that
40\% of gamma rays are picked up and 99.8\% of the background 
cosmic rays are rejected.

\section{Results:}
\label{resul.sec}

We have analyzed about 700 hours observation data of AGNs under the 
good weather condition(corresponds to 2000 hrs $\times$ telescopes exposure).
We have detected clear signals from Mrk501 and Mrk421.
Possible detection of gamma-ray emissions from 1ES1959 is reported in
OG.2.1.21. We could not see clear signals from other sources except for
these three, however, detail analysis is still going on.
In this report, the results of Mrk501 and Mrk421  will be presented.

\begin{figure}[bht]
\begin{center}
\epsfig{file=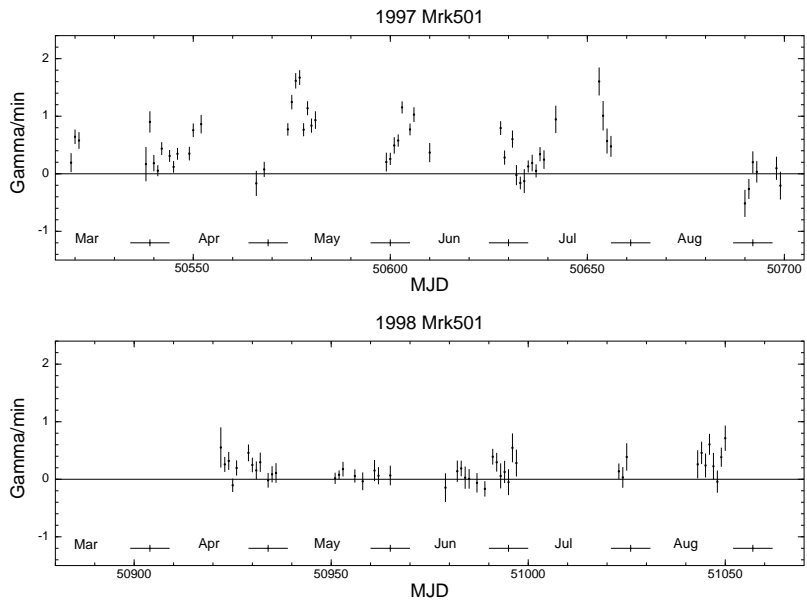,width=4.2in}
\caption{Daily time variation of gamma-ray intensity 
from Mrk501(97) (upper panel) and Mrk501(98) (lower panel).
}
\label{fig:a_tv}
\end{center}
\end{figure}

\begin{figure}[bht]
\begin{minipage}[t]{.47\textwidth}
\begin{center}
\epsfig{file=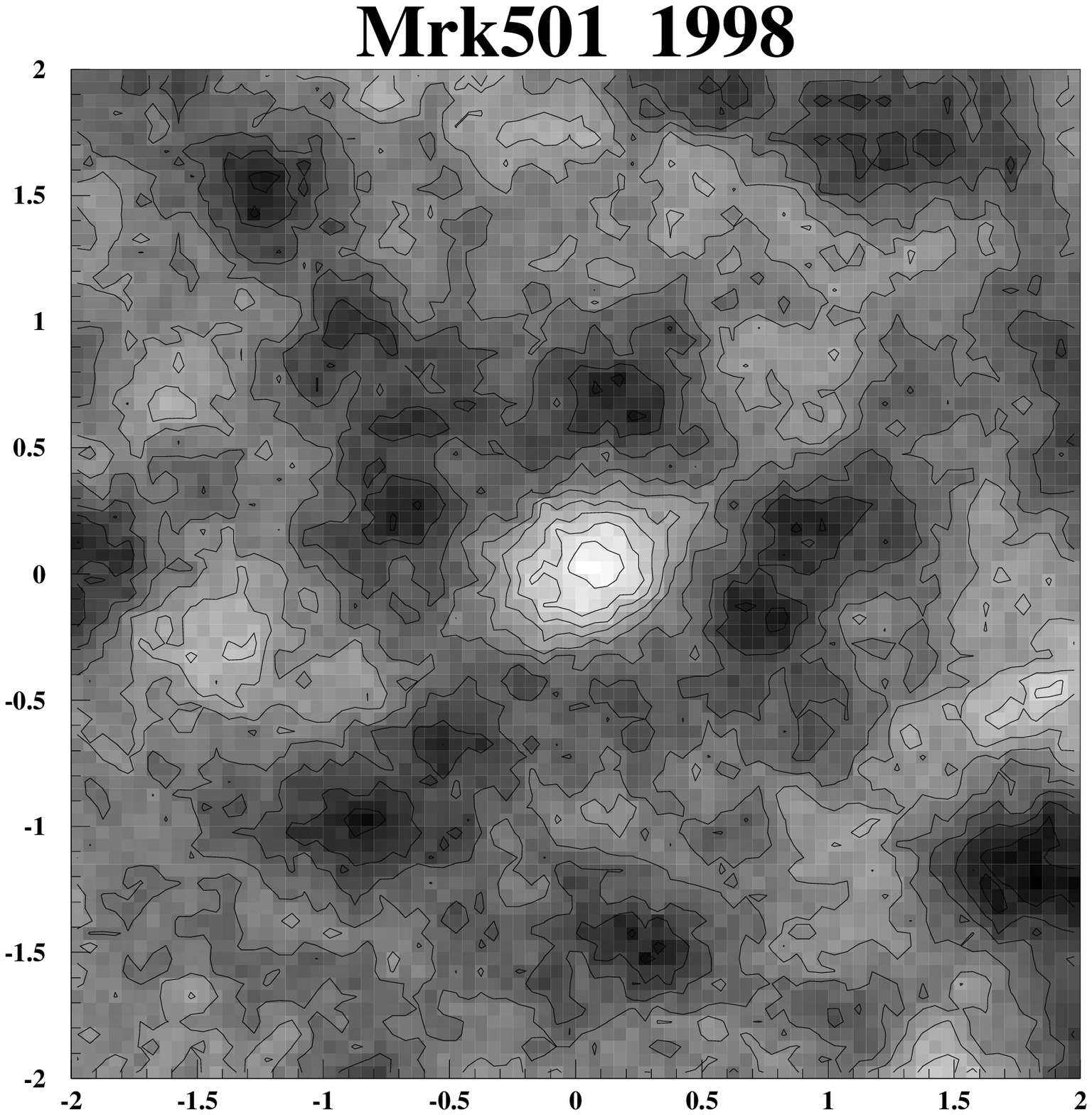,width=2.8in}
\caption{The excess event map of Mrk501(98) in $\alpha < 10^{\circ}$.
Mrk501 is located at the center of the map. The FOV of the map 
is $4^{\circ} \times 4^{\circ}$. }
\label{fig:a_map}
\end{center}
\end{minipage}
\hfill
\begin{minipage}[t]{.47\textwidth}
\begin{center}
\epsfig{file=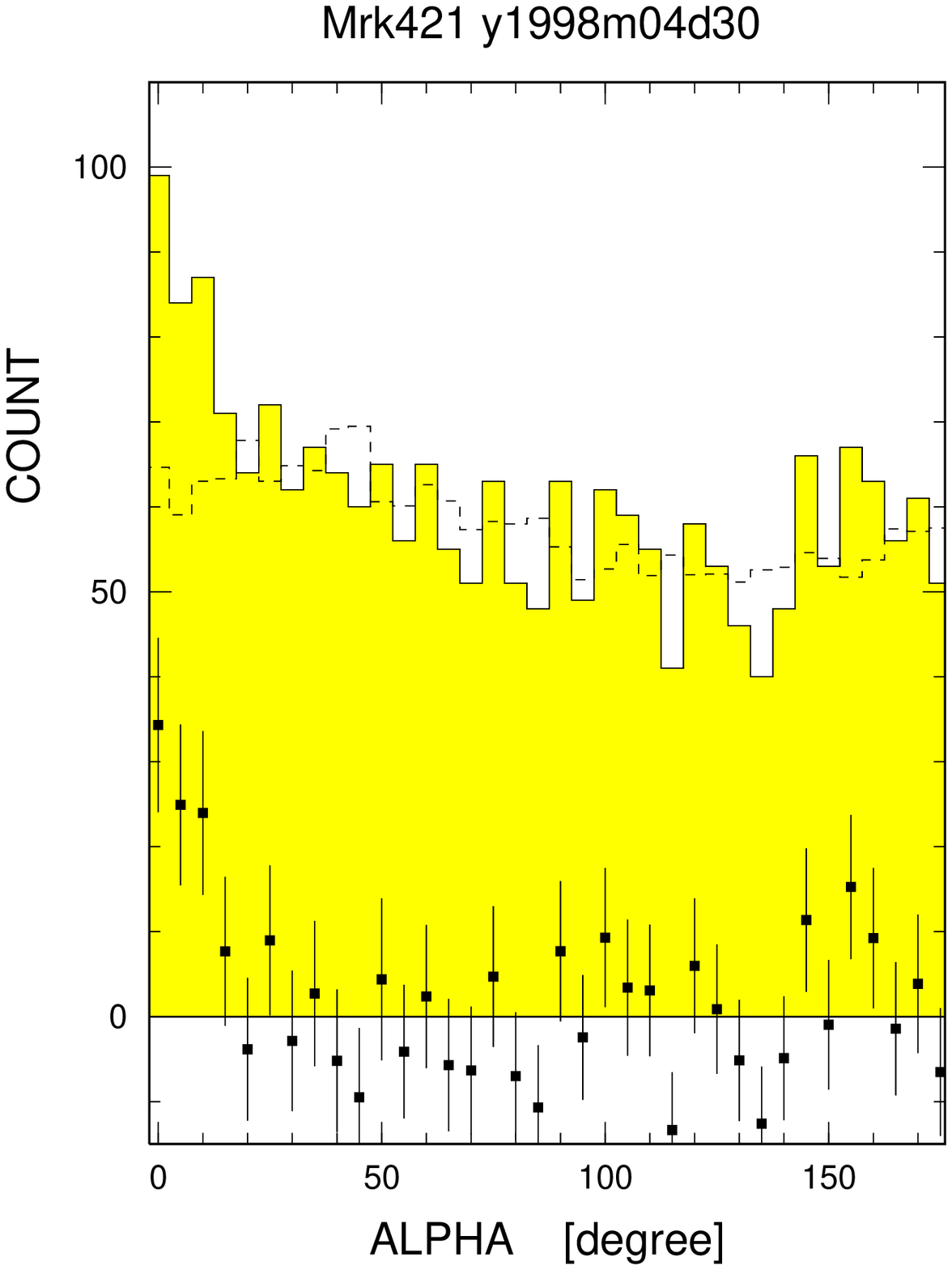,width=2.3in}
\caption{$\alpha$ distribution for observation of Mrk421(30/Apr/98).
Solid line and dashed lines indicate
number of events recorded for on-source and off-source region, respectively.
Dots indicate the number of excess events.}
\label{fig:a7}
\end{center}
\end{minipage}
\end{figure}

Figure \ref{fig:a_tv} shows time variation of gamma-ray intensity 
from Mrk501 in 1997 and 1998. The observation time was 
360 hrs $\times$ tels and
215 hrs $\times$ tels, respectively. Mrk501 was very active in 1997 and
the intensity of the gamma rays varied with about 24 days periodicity
(see OG.2.1.17).
The activity of Mrk501 became low in Sep. 1997.
From the observation in 1998, we could not see any clear flares, however,
most of data points deviate in the positive direction as shown in figure 1.
By adding all data observed in 1998, we obtained 6 sigma excess.
Two dimensional excess map from the Mrk501 for 1998 data is shown in
Figure \ref{fig:a_map}. 
We can conclude that the activity of Mrk501 became lower in Sep.
1997, however, the gamma-ray emission still continued until Sep. 1998.
The Mrk501 should be monitored continuously to investigate the cause 
of the variability of the VHE gamma-ray emission.

The observation of Mrk421 were carried out for 25 nights 
from April to May in 1998, under good weather condition.
Total observation period was 56 hrs.
Among these observations, the most clear excess of 5.3 sigma was found
in 1.3 hrs observation on 30th of April 1998(2450933 MJD). 
$\alpha$ distribution of the data in this flare is shown in Figure \ref{fig:a7}.

\begin{figure}[bht]
\begin{center}
\epsfig{file=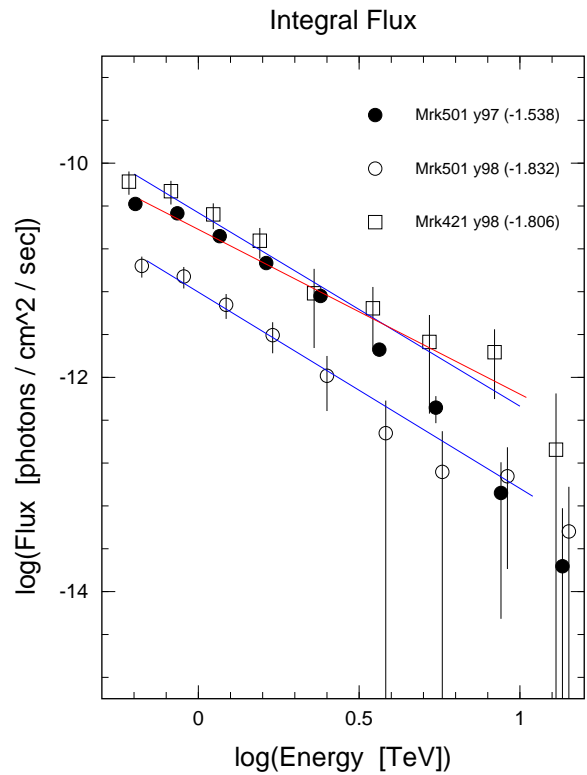,width=3.0in}
\caption{Integral spectrum of Mrk501 and Mrk421.
The best fit of the spectral indices between 900 GeV and 3 TeV are
-1.54, -1.83, and -1.81 for Mrk501(97), Mrk501(98)
and Mrk421(30/Apr/98), respectively.}
\label{fig:a9}
\end{center}
\end{figure}

As reported in the previous paper (Hayashida et al 1998), 
the energy spectrum of Mrk501 in 1997 showed interesting feature,
the spectral index became steeper around several TeV.
If we assume that this lack of higher energy gamma rays is 
due to the interaction with the inter-galactic infrared photons, 
it is expected that the similar
shape appears for the spectrum of Mrk501(98) and Mrk421 because of the 
similar redshifts of Mrk501 and Mrk421.
In Figure \ref{fig:a9}, integral spectra of Mrk421(30/Apr/98), 
Mrk501(97) and Mrk501(98) are compared. 
The indices of the integral spectrum between 900 GeV and 3 TeV
are -1.54 for Mrk501(97), -1.83 for Mrk501(98) 
and -1.81 for Mrk421(30/Apr/98). 
Following this figure, only the spectrum for Mrk501(97) is harder at low energy and 
softer at high energy than those for Mrk421 and Mrk501(98).\\

\hspace{0.5cm}{\bf Acknowledgments}

This work is supported in part by the Grants-in-Aid for Scientific Research 
(Grants \#0724102 and \#08041096) from the Ministry of Education, Science
and Culture. The authors would like to thank the people at Dugway for the 
help of observations.

\vspace{1ex}
\begin{center}
{\Large\bf References}
\end{center}
%
Stecker, F.W., De Jager, O.C.,\& Salamon, M.H., 1996, ApJ 473, L75\\
Hayashida, N. et al, 1998, ApJ 504, L71\\
Nishikawa, D. et al, 1999,Proc.26$^{th}$ ICRC(Salt Lake city,1999)OG.2.1.17\\

\begin{center}
\begin{table}
\begin{center}
\begin{tabular}{|l|c|r|c|l|c|r|}\hline
Object          & z & hours $\times$ && Object & z & hours $\times$  \\
		&   & telescope &&    &   & telescope  \\ \hline
Mrk421          & 0.031 & 561.6  && 1ES0145+138     & 0.125 &  38.0 \\
Mrk501          & 0.034 & 532.1  && EXO0706.1+5913  & 0.125 &  82.9 \\
1ES2344+514     & 0.044 &  43.1  && 1ES0229+200     & 0.139 &  89.0 \\
1ES1959+650     & 0.048 &  250.0 && 1ES0323+022     & 0.147 &  29.9 \\
1ES2321+419     & 0.059 &  101.4 && 1ES0927+500     & 0.188 &  85.5 \\
BL Lac          & 0.069 &  49.6  && 1ES0446+449     & 0.203 &  38.9 \\ \hline
\end{tabular}
\end{center}
\caption{List of AGN survey observation by the Utah Seven Telescope Array. 
	Object names, redshift, observation time 
	which is a product of actual observation time in hour and number of 
	telescopes we used
	under the good weather condition.
	}
\label{tbl:tgt}
\end{table}
\end{center}

\end{document}